\documentclass{ws-rv9x6}
\usepackage{ws-rv-van}     
\makeindex
\newcommand{\GeV}{\,\mathrm{GeV}}
\newcommand{\TeV}{\,\mathrm{TeV}}
\newcommand{\second}{\,\mathrm{sec}}
\newcommand{\be}{\begin{equation}}
\newcommand{\ee}{\end{equation}}
\newcommand{\ba}{\begin{array}}
\newcommand{\ea}{\end{array}}
\newcommand{\bea}{\begin{eqnarray}}
\newcommand{\eea}{\end{eqnarray}}
\newcommand{\eq}[1]{Eq.~(\ref{eq:#1})}
\begin{document}

\chapter[The string landscape and low energy supersymmetry]%
{The string landscape\\ {\it and}\\ low energy supersymmetry}
\label{ra_ch1}

\author[M. R. Douglas]{Michael R. Douglas}

\address{Simons Center for Geometry and Physics\\
Stony Brook University\\
Stony Brook, NY 11794 USA\\
{\tt mdouglas@scgp.stonybrook.edu}}

\begin{abstract}
We briefly survey our present understanding of the string landscape, and use it to discuss the chances that
we will see low energy supersymmetry at the LHC.\footnote{
This is a revised version of a contribution to 
{\it Strings, Gauge Fields, and the Geometry Behind - The Legacy of Maximilian Kreuzer,}
to be published by World Scientific.}
\end{abstract}

\body

\section{The goal of fundamental physics}\label{ra_sec1}

Max Kreuzer will be remembered as one of the true pioneers of string theory.
His untimely passing is a great loss for his family, for his many friends, and for science.
Although I did not know him as well as many of you,
I was fortunate to have the chance to co-organize a workshop with him
here at the ESI in 2008, and I am honored to join you today in celebrating his
many contributions to science.

Among these contributions, one which greatly influenced me, and which has been admired
by many of the speakers here, is his survey with Skarke of the three dimensional Calabi-Yau toric hypersurfaces.
This will be remembered as the first picture of the string landscape.  While there were glimmerings of its existence from
other arguments, based on the cosmological constant problem, or generic properties of constructions
involving a large number of combinatorial choices,
this work was based on a key aspect of string theory compactification which has
not been transformed by subsequent developments.

Particle physics is now at a turning point.  The LHC has been operating for over a year now at $7  \TeV$, and many 
results have been announced.  In December 2011, just before this was written, significant (though not conclusive)
evidence for a Higgs boson at $125 \GeV$ was announced by both ATLAS and CMS.  On the other hand, there has
been no evidence for any non-Standard Model physics.  In particular, colored gauginos, a signature of low
energy supersymmetry, have been excluded below about $500 \GeV$.  Over the coming year,
several times as much data (over 10 ${\rm fb}^{-1}$) will be accumulated, realizing most of the discovery potential for runs
at this energy.  Afterwards, while supersymmetry (or other non-SM physics) might still be discovered at LHC, 
this will require either the
upgrade to $14 \TeV$, or many years of data taking and subtle analysis to uncover superpartners which only have
electroweak interactions, or perhaps both.

By now it is almost a truism that string theory makes no definite predictions for LHC physics,
only suggestions for rather implausible scenarios such as black hole creation, whose non-discovery
would not falsify the theory.  This is not literally true as there are potential discoveries which would give
strong evidence against string theory,\footnote{
This includes time-varying $\alpha$ \cite{Banks:2001qc}, and probably faster-than-light neutrinos.}
but at present there is no reason to expect them.

Even if we find no ``smoking gun'' which speaks directly for or against the theory, there is a program which could
someday lead to falsifiable predictions.  It is to understand the landscape of string vacua, and derive a probability
measure on the set of vacua based on quantum cosmology.
From this, we can infer the probabilities that each of the various possibilities for beyond
the Standard Model, cosmological, and other fundamental physics would come out of string theory.  If future
discoveries, and (to some extent) present data come out as highly unlikely by this measure, we have evidence against
string theory under the assumed scenario for quantum cosmology.   This evidence might or might not be conclusive, but it would
be the best we could do with the information to hand.

This program is only being pursued by a few groups today and would require major advances in our understanding
of both string theory and quantum cosmology before convincing predictions could be made.  My guess at present
is that twenty years or more will be needed, taking us beyond the LHC era.  Even then, it is likely that
such predictions would depend on hypotheses about quantum cosmology which could not be directly tested and might admit
alternatives.  It is entirely reasonable that sceptics of the landscape should reject this entire direction and look for 
other ways to understand string theory, or for other theories of quantum gravity.  At present we do not know enough
to be confident that they are wrong.  Nevertheless the evidence at hand leads me to think that they are wrong and that this
difficult path must be explored.

In this short note, I will briefly outline this program and how I see it proceeding.
Although it is clearly a long term project, after today's sad reminder that our individual existences are
limited and that each of us
must still try to see as far as he or she can, I am going to go out on a limb and argue that
\begin{quotation}
String/M theory will predict that our universe has supersymmetry, broken at the $30-100 \TeV$ scale.
If at the lower values, we may see gluinos at LHC, while if at the higher values,
it will be very hard to see any evidence for supersymmetry.
\end{quotation}
This is a somewhat pessimistic claim which far outruns our ability to actually make predictions from
string theory.
Nevertheless I am going to set out the argument, fully realizing that many of the assumptions
as well as the supporting evidence might not stand the test of time.  Indeed, we should all hope that this is wrong!

To begin, we have to make the case that a fundamental theory should allow us to make any predictions of this
scope.  This is not at all obvious.  Certainly most major scientific discoveries were not anticipated
in any detail.  However the record in particle physics is far better, with examples including the positron, neutrinos,
the charm quark, the third generation of quarks and leptons,
the W and Z bosons, and as it now appears, the Higgs boson.  
The framework of quantum
field theory is highly constraining, and this record of success is the evidence.

Of course, quantum field theory is only constraining within certain limits.  
For example, there is no good argument which
favors three generations of quarks and leptons over four.  There are many other consistent
extensions of the Standard Model which we might imagine discovering.  Even the basic structure
of the Standard Model, its gauge group and matter representation content, admits consistent variations.
As things stand, it is entirely reasonable to claim that this structure was a choice which could not have been
predicted {\it a priori}, and equally that the existence of as yet undiscovered matter cannot be excluded
{\it a priori}.

Quantum gravity is hoped to be more constraining, although there is no consensus yet on whether or why this
is true.  In the case of string/M theory,  we can benefit from over 25 years of work on string compactifcation.
It is clear that there are several different constructions based on the different perturbative limits and compactification
manifolds: heterotic on a Calabi-Yau manifold, F theory, type II with branes, M theory on $G_2$ manifolds.
Each involves many choices which lead to different outcomes for low energy physics, and for this reason one
cannot make definite predictions.  

There is even an extreme
point of view that (in some still vague sense) ``all''  consistent low energy theories can be realized as string
compactifications.  This idea can lead in various directions: one can hope that consistency will turn out to be a more powerful
constraint in quantum gravity than it was in quantum field theory.\cite{ArkaniHamed:2006dz}
On the other hand, this does not seem to be the case in six space-time dimensions.\cite{Kumar:2009us}
And so far, in four dimensions,  no-go results are surprisingly rare.

While there are too many compactifications to study each one individually,
one can hope that a particular construction or class of compactification would lead to some
generic predictions.  If the broad structure of the SM or some BSM scenario came out this way, one might hypothesize
that that construction was preferred, and look for top-down explanations for this.  However no construction seems
especially preferred at this point.  For example, it is simple to get grand unification out of the $E_8\times E_8$ heterotic
string.  On the other hand, this construction does not naturally lead to three generations of matter; even if we grant
that the number of generations must be consistent with asymptotic freedom, most choices of Calabi-Yau manifold
and bundle have other numbers of generations.  Can one do better?
There are brane constructions which relate this `three' to the
number of extra (complex) dimensions, but these do not realize grand unification.  Which is better?

It seems that any attempt to narrow down the possibilities will involve this type of weighing of different factors,
and this is a strong motivation to systematize this weighing and make it more objective.  While this would seem like
a very open-ended problem, in the context of string compactification there is a natural way to do it -- namely, to
count the vacua of different types, and regard high multiplicity vacua as favored or ``more natural.''  This includes
the considerations of tuning made in traditional naturalness arguments, and extends them to discrete and
even qualitative features such as numbers of generations or comparison of different supersymmetry breaking mechanisms.
The basic outlines of such an approach are set out in Ref.~\refcite{Douglas:2006es}, and have led so far to a few 
general results which we will survey below.
One can imagine continuing this study along formal, top-down lines to develop a quantitative picture of the landscape.

While this sort of information seems necessary to proceed further, by itself it is not going to lead to convincing
predictions.  I like the analogy to the study of solutions of the Schr\"odinger equation governing electrons
and nuclei, better known as chemistry.\cite{Douglas:2006za}  
The landscape of chemical molecules is very complicated, but one could
imagine deducing it {\it ab initio} and working out a list of long-lived metastable compounds and their properties.
However, in any real world situation (both on earth, and in astronomy), the number density of the various molecules
is very far from uniform, or from being a Boltzmann distribution.  One needs some information about the processes which
created the local environment, be it the surface of the earth, the interior of a star, or whatever, to make any
{\it ab initio} estimate of this number density.  Conversely, we see in astrophysics that fairly simple
models can sometimes lead to useful estimates.  One can then make statements about ``typical molecules,''
meaning typical for that local environment, on purely theoretical grounds.

While one cannot push this analogy very far, I think it confirms the point that we need some information about the
processes which created our vacuum as one of the many possibilities within the landscape, to estimate a measure
and make believable predictions.  Doing this is a primary goal of quantum cosmology and has been discussed for over 30 years.
The first question is whether one needs the microscopic details of quantum gravity to do this, or whether general
features of quantum gravity suffice.  There is a strong argument, based on the phenomenon of eternal inflation,
that the latter is true, so that one can ask the relevant questions and set up a framework to answer them without
having a microscopic formulation.  Of course, their answers might depend on microscopic details; for example 
one needs to know which pairs of vacua are connected by tunnelling processes, and this depends on the structure of
configuration space.  Anyways, these general arguments are well reviewed in 
Refs.~\refcite{Guth:2007ng,Linde:2007zz,Vilenkin:2011dd}.

Quantum cosmology is a contentious subject in which I am not an expert.  However, within the 
eternal inflation paradigm, starting from a variety of precise definitions for the measure factor
and using the presence of many exponentially small numbers in the problem, one obtains
a fairly simple working definition, the ``master'' or ``dominant'' vacuum 
ansatz.\cite{Freivogel:2011eg,SchwartzPerlov:2006hi,Harlow:2011az}
This states that the {\it a priori} measure is overwhelmingly dominated by the longest lived metastable de Sitter vacuum.
The measure for other vacua is given by the tunnelling rate from this ``master'' vacuum,
which to a good approximation is that of the single fastest chain of tunnelling events.

Although the measure is dominated by the master vacuum, it is {\it a priori} likely (and we will argue) that observers
cannot exist in this vacuum -- it is not ``anthropically allowed.''   While there are many objections to anthropic 
postselection, they have been well addressed in the literature, and we will not discuss them here.  
Nevertheless
we must take a position on what anthropic postselection should mean in practice.  The philosophically correct
definition that a vacuum admit observers is impossible to work with, while simpler proxies such as entropy 
production\cite{Bousso:2010vi} have not yet been developed in the detail we need for particle physics.  

In practice, the anthropically allowed vacua will be those which realize the Standard Model gauge group,
and the first family of quarks and leptons, with parameters roughly the ones we observe.
It is not obvious that even these are all anthropically selected; for example Ref.~\refcite{Harnik:2006vj} argues
that one does not even need the weak interactions!\footnote{
Even granting this point, the need to get several quarks and leptons with similar small nonzero masses is 
far more easily met by a chiral theory such as the Standard Model, than a vector-like weakless theory.}  
Conversely, while the precise values of quark
masses are not usually considered to be selected, given the plethora of
fine tunings in chemistry, 
it might well be that life and the existence of observers is much more dependent on the specific
values of these parameters than it first appears.

Besides these questions of detail, any definition of anthropic selection suffers from the objection that it is time
dependent and would be different in 1912 or 2112 than in 2012.  While this is so, we would reply that all we can do in
the end is to test competing theories with the evidence to hand, and one can try out all the variations on this theme
in order to do this.  It is quite reasonable to expect our evidence to improve with time, and perhaps our understanding
of the anthropic constraints will improve as well.

Granting that the master vacuum is not anthropically allowed, the measure we are interested in is thus the ``distance''
in this precise sense (defined using tunnelling rates)
from the master vacuum, restricted to the anthropically allowed vacua.  Clearly it is important to
find the master vacuum, and one might jump to the conclusion that
string theory predicts that we live in a vacuum similar to the master vacuum.  However, because of the anthropic
constraint, whether this is so depends on details of the tunnelling rates.  The main constraint is that one needs to 
reach a large enough set of vacua to solve the cosmological constant problem.  The more tunnelling events required to
do this, and the more distinct the vacua they connect, the more disparate a set of vacua will fall into this category,
and the weaker the predictions such an analysis would lead to.  Somewhat counterintuitively, if the master vacuum admits
many discrete variations, then there are more nearby vacua and one does not need to go so far to solve the cosmological 
constant problem.  If this set of vacua includes anthropically allowed vacua, then these would be favored and one can
imagine getting fairly definite predictions.

One can already make some guesses for where to look for the master vacuum in the string landscape, as we will describe.
Continuing in this speculative vein, we will argue that this favors ``local models'' of the SM degrees of freedom, and
supersymmetry breaking driven by dynamics elsewhere in the extra dimensions, gravitationally mediated to the SM.
This is a much-discussed class of models and, as we discuss in section 3,
a key problem which limits their ability to solve the hierarchy problem is the cosmological moduli problem,
which seems to require supersymmetry breaking at or above around $30 \TeV$. 
Still, compared to the GUT or Planck scales, this is a huge advantage, and thus
we predict low energy susy but with superpartners around this scale.

The argument we just gave is not purely top-down and is closely related to the familiar arguments
that if low energy supersymmetry were the solution to the hierarchy problem, we should see superpartners in the current
LHC runs.   My phenomenology is rather sketchy and there are many other scenarios that would need to be considered
to make a convincing argument.  But the point here is to illustrate
the claim that with some additional input from the string theory landscape, allowing
us to compare the relative likelihood of different tuned features, we could make such arguments precise.

\section{Low energy supersymmetry and current constraints}

Most arguments for ``beyond the Standard Model'' physics are based on its potential for solving the
hierarchy problem, the large ratio between the electroweak scale $M_{EW} \sim 100 \GeV$ and higher
scales such as $M_{Planck}\sim 10^{19} \GeV$ or $M_{GUT} \sim 10^{16} \GeV$.  Low energy supersymmetry
is a much-studied scenario with various circumstantial arguments in its favor.  Theoretically, it is
highly constraining and leads to many generic predictions, most importantly the gauge couplings
of superpartners.  This is why LHC already gives us strong lower bounds on the masses of colored superpartners,
especially the gluino. 

If there is a $125 \GeV$ Higgs, this turns out to put interesting constraints on supersymmetric models.
Recent discussions of this 
include Refs.~\refcite{Akula:2011aa,Baer:2011ab,Hall:2011aa}, the talk at Ref. \refcite{ArkaniHamed:2012},
and Ref. \refcite{Acharya:2012tw} which reviews a line of work which influenced my thoughts on these questions.

A broad brush analysis of the hierarchy problem can be found in Ref.~\refcite{Barbieri:2003dd}.
Its solution by low energy supersymmetry can be understood by restricting
attention to a few fields, most importantly the top quark and its scalar partner the `stop'.  
In general terms, top quark loops give a quadratically divergent contribution to the Higgs mass, which is
cut off by stop loops.  This leads to the rough estimate
\be \label{eq:higgs-est}
\delta M_{H1}^2 \sim  0.15 M_{ST}^2 \log \frac{\Lambda_{SUSY}}{M_{ST}} ,
\ee
where $M_{H1}$ is the mass of the Higgs which couples to the top,
$\Lambda_{SUSY} \equiv M_{3/2}$ is the supersymmetry breaking scale 
and $M_{ST}^2$ is the average stop mass squared.

The strongest sense in which supersymmetry could solve the hierarchy problem would be to ask, not just
that $M_H$ comes out small in our vacuum, but that it comes out small in a wide variety of vacua similar
to ours; in other words all contributions to $M_H$ are of the same order so that no fine tuning is required.
This is called a natural solution and from \eq{higgs-est} it requires
$$
M_{ST} \sim M_{H} 
$$
in a fairly strong sense (for example Ref. \refcite{Brust:2011tb} estimates $M_{ST} \lesssim 400 \GeV$),
Although this might sound like it is already ruled out, the stop cross section is quite a bit smaller than
that of the gluino, and there are even scenarios in which the lightest stop is hard to find because it is
nearly degenerate with the top.

While $\Lambda_{SUSY} \sim M_{EW}$ as well, 
there are many different types of supersymmetry breaking and this does not in itself require the gluino to be light.
But one can get a much stronger constraint by assuming that $M_{ST}$ is naturally low as well, as it gets mass 
renormalization from gluon and gluino loops.  Assuming that there are no other colored particles involved,
this leads to an upper bound on the gluino mass \cite{Brust:2011tb}
$$
M_{\tilde g} \lesssim 2 M_{ST} .
$$
Thus, LHC appears to be on the verge of ruling out a wide variety of natural models similar to the MSSM.

These low bounds on the masses of superpartners in natural models were already problematic before LHC for a variety of
reasons, but most importantly because of the difficulty of matching precision measurements in the Standard
Model.   The longest standing problem here is the absence of flavor changing processes other than those
mediated directly by the weak interactions, which translates into lower bounds for the scale of much new
physics of $\Lambda \gtrsim 10-100 \TeV$ !  

Assuming the superpartners are not found below $1 \TeV$, a reasonable response is to give up on naturalness and
accept some tuning of the Higgs mass.
The cleanest such scenario is to grant the standard structure of low energy supersymmetry, but push
it all up to the $10-100 \TeV$ scale.  Thus, all of the superpartners and the Higgs bosons would {\it a priori}
lie in the range $0.1 - 1$ times $\Lambda_{SUSY}$, but we then postulate an additional $10^{-4}-10^{-6}$ fine tuning of 
one of the Higgs boson masses.  
At first sight this has the problem that we lose the WIMP as a candidate dark matter particle. 
However, because  the gauginos have R charge and the scalars do not,
it is natural for them to get a lower mass after supersymmetry breaking.  

In a bit more detail, one very generally expects irrelevant interactions between the supersymmetry breaking
sector and the Standard Model sector to produce soft masses for all the scalars of order $\Lambda_{SUSY}= M_{3/2}$.
In the original supergravity models, this scale was set to $M_{EW}$, but this leads to many lighter particles and
by now has been ruled out.  One way to try to fix this is gauge mediation, in which other interactions provide
larger soft masses.  As far as the SM is concerned this may be good, but it suffers from the cosmological
moduli and gravitino problems we will discuss in \S \ref{ss:modprob}.
One can instead try to work with  $\Lambda_{SUSY} \gg M_{EW}$, and argue that the naive expectations
for the soft masses are incorrect.  There are many ideas for this, such as sequestering\cite{Randall:1998uk}
(see Ref. \refcite{Berg:2010ha} for a recent string theory discussion), focus point models,\cite{Feng:1999zg} 
intersection point models,\cite{Feldman:2011ud} and others.  Clearly it would be important if a generic
mechanism could be found, but as yet none of these have found general acceptance.  Thus we accept the generic
result $M_0\sim\Lambda_{SUSY}$ for scalar masses.  However,
the gaugino soft masses have other sources and, as we will discuss below, can be smaller.

The extreme version of this
scenario is split supersymmetry,\cite{ArkaniHamed:2004fb,ArkaniHamed:2004yi} in which the scalars
can be arbitrarily heavy while all fermionic superpartners are light.  In any case, although $M_H$ is tuned,
it is essentially determined by the quartic Higgs coupling and the Higgs vev, which we know from $M_Z$.
If the underlying model is the MSSM, then since the quartic Higgs coupling comes from a D term, 
it is determined by the gauge couplings
and we get a fairly direct prediction for the Higgs mass.  As is well known, at
tree level there is a bound $M_H^2 \le M_Z^2$, and one must call on \eq{higgs-est} just to satisfy
the LEP bound $M_H > 113 \GeV$.   There has been much recent discussion of the difficulty of getting
$M_H \sim 125 \GeV$ to come out of the MSSM in a natural way.%
\cite{Akula:2011aa,ArkaniHamed:2012,Baer:2011ab,Hall:2011aa}

If $M_H$ is fine tuned, the loop contribution \eq{higgs-est} is no longer directly measurable.  However there
is a similar-looking constraint coming from the running of the  quartic Higgs coupling, of the general form\footnote{
This is a bit simpified, and the precise expression depends on ones' assumptions, 
see for example Refs. \refcite{Baer:2011ab,ArkaniHamed:2004fb}.  The main difference with \eq{higgs-est}
is that one has put in the observed electroweak parameters and thus accepted the possibility of tuning.}
\be \label{eq:higgs-two}
M_{H}^2 \sim M_Z^2 \cos^2 2\beta
 + \frac{3 g^2 M_{top}^4}{16\pi^2 \sin^2\beta M_W^2} \log \frac{\Lambda_{SUSY}}{M_{top}} .
\ee
In fact, for $M_H=125 \GeV$, this turns out to predict $\Lambda_{SUSY} \sim 100 \TeV$.
This prediction is not robust; for example by postulating another scalar which couples to the Higgses through the
superpotential (the NMSSM), or an extra $U(1)$,
one can change the relation between the Higgs quartic coupling 
and the gauge coupling.  Such modifications will affect the Higgs branching ratios, so this alternative 
will be tested at LHC in the coming years.

Within this scenario, the key question for LHC physics is whether we will see the gauginos.  There are many
models, such as anomaly mediation,\cite{Randall:1998uk,Giudice:1998xp,Bagger:1999rd} 
in which the expected gaugino mass is
\be
M_\chi \sim \frac{\beta(g)}{2g^2} \Lambda_{SUSY} .
\ee
where $\beta(g)$ is the exact beta function.
This prefactor is of order $10^{-2}$ for the neutralinos, so we again have a dark matter candidate if
$\Lambda_{SUSY} \sim 100 \TeV$.
For the gluino, it is $\sim 1/40$  and thus we are right at the edge of detection at LHC-14.  
With $\Lambda_{SUSY} \sim 30 \TeV$ we would have gluino mass
$M_3 \sim 750 \GeV$ which should be seen very soon.

Note that there are other string compactifications with large gaugino masses.\footnote{I thank Bobby Acharya,
Gordy Kane and Gary Shiu for discussions on this point.}
The relevant contribution is $F^a \partial_a f$ where $F^a$ is an $F$-term and $f$ is a gauge kinetic term.
Thus, gaugino masses will be large if the supersymmetry breaking F terms are in fields, such as the heterotic
string dilaton, whose expectation
values strongly affect the observed gauge couplings.  This is a question about the supersymmetry breaking
sector which must be addressed top-down; in section 6 we will suggest that small masses are preferred.

It would be
very valuable from this point of view to know the lower limit on $\Lambda_{SUSY}$ in these scenarios,
as all other things being equal, by naturalness it seems reasonable to expect the lower limit (we will discuss 
stringy naturalness later).

\section{The gravitino and moduli problems}
\label{ss:modprob}

It has been known for a long time that light, weakly coupled scalars can be problematic for inflationary 
cosmology.\cite{Coughlan:1983ci}  Because of quantum fluctuations, on the exit of inflation they
start out displaced from any minimum of the potential, leading to the possible ``overshoot'' of the desired
minimum, and to excess entropy and/or energy.  

In string/M theory compactification, moduli of the extra dimensional metric and other fields very generally lead after
supersymmetry breaking to scalar fields with gravitational strength coupling and mass $M\sim \Lambda_{SUSY}$,
making this problem very relevant.\cite{Banks:1993en,deCarlos:1993jw}
For $M \sim 1 \TeV$, such particles will decay around  $T \sim M_{Planck}^2/M^3 \sim 10^3 \second$.
This is very bad as it spoils the predictions for abundances of light nuclei based on big bang nucleosynthesis,
which takes place during the period $0.1 \lesssim T \lesssim 100 \second$.  One needs such particles to be either
much lighter, or much heavier, so that they decay at $T << 0.1 \second$.  One can also increase their couplings
so that their decay reheats the universe well above the scale of nucleosynthesis -- see Ref. \refcite{Acharya:2008bk}
for a recent proposal of this type.  There is a closely related constraint from gravitino decay,\cite{Ellis:1984eq}
which also can be solved this way.

Although not proven, it is quite generally stated that this constraint forces the moduli masses to satisfy
\be \label{eq:moduli-bound}
M_{moduli} \gtrsim 30 \TeV .
\ee
This may not {\it a priori} require $\Lambda_{SUSY} \gtrsim 30 \TeV$, as there are other ways to lift
moduli masses.  For example, fluxes can give masses to moduli while preserving 
supersymmetry.\cite{Giddings:2001yu,Douglas:2006es}  On the other hand, the structure of the 
$N=1$ supergravity potential makes it generic to have at least one scalar with $M \lesssim M_{3/2}$,
as shown in Refs.  \refcite{Denef:2004cf,Acharya:2010af}.  This is a scalar partner of the goldstino and
it need not have gravitational strength interactions, but if supersymmetry breaking takes place in a
hidden sector this will often lead to a constraint.  In addition, the gravitino has $M=M_{3/2}$ by
definition.  If one is going to call on reheating or other physics to solve these problems, it is simplest
and probably most generic for the problematic particles to be associated with a single energy scale.

These various considerations all suggest that 
\be\label{eq:susy-bound}
\Lambda_{SUSY} = M_{3/2} \gtrsim 30 \TeV .
\ee
This is an extremely strong constraint which very much disfavors the natural solutions to the hierarchy
problem, and is independent of the arguments we gave in the previous section.  This is widely recognized
and thus there has been a major effort to search for generic ways out or at least loopholes to this bound.
On the other hand, even if there are loopholes, the bound still might be generic.  In the context of the string
landscape, the correct attitude would then be to accept it as preferred, also allowing the vacua (and cosmological
histories) which realize the loophole, but weighing them by appropriate tuning factors.  We will begin
this discussion below.

Perhaps the main reason that \eq{susy-bound} has not been more widely accepted is that
there is no direct evidence at this point for the significance of the energy scale $30-100 \TeV$.
This is of course related to the lack of direct evidence for supersymmetry, so perhaps we should not
be too bothered by this, but we should ask for some more fundamental reason why 
$\Lambda_{SUSY} \sim 30-100 \TeV$ should be preferred.  Later we are going to argue this from
stringy naturalness, essentially that $\Lambda_{SUSY}$ should take the lowest possible value consistent
with anthropic constraints.  

A weaker but more general claim, less dependent on anthropic constraints, is that the cosmological moduli problem
will always favor a ``little hierarchy'' between the mass scale of ``normal'' matter
and the supersymmetry breaking scale.  Let us start from a more general statement of the problem -- it
is that important cosmological physics (in our universe, BBN) takes place at a temperature just below the mass 
of normal matter, and thus moduli (and the gravitino) must decay well before this happens and/or reheat the
universe above this temperature.  Thus, we have
\be
T_{reheat} > c M_{matter}
\ee
with a constant $c \sim 10^{-3} - 10^{-1}$.  Given $T_{reheat} \sim \Lambda_{SUSY}/M_{Pl}^{1/2}$, this implies
\be
\Lambda_{SUSY}^3 > c M_{Planck} M_{matter}^2
\ee
and the large hierarchy $M_{Planck} \gg M_{matter}$ forces $\Lambda_{SUSY} \gg M_{matter}$, but
only as the one-third power of $M_{Planck}/M_{matter}$ and suppressed by the constant $c$.

An even more broad brush way of arguing would be to say that inflationary cosmology is already difficult enough to make 
work at each of the relevant scales (the matter/BBN scale, the electroweak scale and now the supersymmetry
breaking scale) that one should expect at least little hierarchies between these various scales just to simplify the
problem.  Whether this simplicity is of the type that Occam would have favored, or whether it has any relevance
for stringy naturalness, remains unclear.

\section{The set of string vacua}

The broad features of string compactification are described in Ref.~\refcite{Douglas:2006es} and many other reviews.
We start with a choice of string theory or M theory, of compactification manifold, and of the topological class
of additional features such as branes, orientifolds and fluxes.  We then argue that the corresponding supergravity
or string/M theory equations have solutions, by combining mathematical existence theorems ({\it e.g.} for a
Ricci flat metric on a Calabi-Yau manifold), perturbative and semiclassical computations of corrections to
supergravity, and general arguments about the structure of four dimensional effective field theory.

Perhaps the most fundamental distinction is whether we grant that our vacuum breaks $N=1$ supersymmetry
at the compactification (or ``high'')
scale, or whether we can think of it as described by a four dimensional $N=1$ supersymmetric effective
field theory, with supersymmetry breaking at a lower scale.  Almost all work makes the second assumption, largely because
there are no effective techniques to control the more general problem, nor is there independent evidence 
(say from duality arguments) that many high scale vacua exist. 
Early work suggesting that such vacua were simply the large $\Lambda_{SUSY}$ limit of the usual supersymmetric 
vacua\cite{Douglas:2004qg,Susskind:2004uv}
was quickly refuted by a more careful analysis of supersymmetry breaking.\cite{Denef:2004cf}

A heuristic and probably correct argument that this type of nonsupersymmetric vacuum is
very rare is that stability is very difficult to achieve
without supersymmetry -- recently this has been shown in a precise sense for random supergravity 
potentials.\cite{Marsh:2011aa}
There are many versions of this question, some analogous and some dual, such as the
existence of Ricci flat metrics without special holonomy, and the existence of interacting conformal field theories
without supersymmetry.  We will assume that metastable nonsupersymmetric vacua are not common enough to 
outweigh their disadvantages;
of course, if a large set of them were to be discovered, this would 
further weaken the case for low energy supersymmetry.

Another context in which nonsupersymmetric vacua might be very important is for the theory of inflation.
A natural guess for the scale of observed inflation is the GUT scale, in other words the compactification scale.
The requirement of near-stability is still very constraining, however, and almost all work on this problem
assumes broken $N=1$ supersymmetry as well.  Since the inflationary trajectory must end up in a metastable
vacuum, it is hard to see how it could be very different from this vacuum anyways.

Granting the need for 4d $N=1$ ``low scale'' supersymmetry (here meaning compared to the string theoretic scales),
each of the five 10d string theories as well as 11d supergravity have a preferred extra dimensional geometry
which leads there.   Some theories (such as type I and $SO(32)$ heterotic) disfavor the Standard Model, 
but other extra dimensional interpretations (such as F theory) were developed, leading to this table:\break
\begin{table}[h]
\begin{tabular}{ccc}
\hline \\
heterotic M theory & M theory & F theory \\
(CY threefold) & (G$_2$ manifold) & (CY fourfold) \\
\vdots & \vdots & \vdots\\
$E_8\times E_8$ string & IIa string with D6 & IIb string with D3, D7 \\
\\
\hline
\end{tabular}
\end{table}

\noindent
The arrangement reflects the duality relations between the theories, with the vertical axis corresponding to
adding an extra dimension, while the horizontal axis allows various dualities depending on the fibration
structure of the manifolds involved (heterotic-IIa, mirror symmetry, and others).  

Within each of these constructions, one can make fairly concrete pictures of the sources of gauge symmetry,
matter and the various interactions, as arising from higher dimensional gauge fields and their fermionic
partners (possibly living on branes),
wave function overlaps or brane intersections, and instantons.   These lead to generic predictions such as the
presence or absence of grand unification and certain matter representations, but in general there is a lot
of freedom to realize the Standard Model and a wide variety of additional matter sectors.

An important distinction can be made between ``global'' models such as
heterotic string compactification, and ``local'' models such as F theory.   
In a global model, realizing
chiral matter requires postulating structure on the entire extra dimensional manifold.  
By contrast, in a local model, chiral matter can be realized at the intersection of branes which are contained
in some arbitrarily small subregion of the manifold.  This is nontrivial because chiral matter can only be
realized by brane intersections which (in a certain topological sense) span all of the extra 
dimensions.\cite{Berkooz:1996km}  While naively this makes local models impossible, and on simple
topologies such as an $n$-torus they would be impossible, they are possible
in more complicated geometries such as resolved orbifolds and elliptic fibrations.

Local models tend not to realize gauge unification, and in the simplest examples cannot realize the
matter representations required for a GUT, such as the spinor of $SO(10)$.  These two problems were
more or less overcome by the development of F theory local models.\cite{Beasley:2008dc,Donagi:2008ca}
F theory is also attractive in that one can more easily understand the other constructions by starting from
F theory and applying dualities, than the other way around.

It was suggested in Ref. \refcite{Beasley:2008dc} that local models should be preferred because
they admit a consistent decoupling limit.  Essentially, this is a limit in which one takes the small subregion containing
the local model to become arbitrarily small.  Because observable scales (the Planck scale and the scale of matter)
tend to be related to scales in the extra dimensions, it is more natural to get hierarchies in this limit.  At present
the status of this argument is extremely unclear, as it is generally agreed that global models can realize hierarchies
through dynamical supersymmetry breaking and otherwise.  Later we will discuss a different, cosmological argument
that might favor local models.

Because of the dualities, and the existence of topology changing transitions in string/M theory,
the usual picture is of a single ``configuration space'' containing all the vacua and allowing transitions (perhaps
via chains of elementary transitions) between any pair of vacua.  Only special cases of this picture have been worked out; 
for example it has long been known that all of the simply connected Calabi-Yau threefolds are connected by 
conifold transitions.  More recently, ``hyperconifold transitions'' were introduced
which can change the fundamental group,\cite{Davies:2009ub,Davies:2011is} 
but it is not known whether these connect all the
non-simply connected threefolds.

It is very important to complete this picture and develop concrete
ways to represent and work with the totality of this configuration space.  Even its most basic properties, such as any sense
in which it is finite, are not really understood.  Various ideas from mathematics can be helpful here; in particular there
is a theory of spaces of Riemannian manifolds in which finiteness properties can be proven, such as Gromov-Cheeger
compactness.  Very roughly, this says that if we place a few natural restrictions on the manifolds, such as an
upper bound on the diameter (the maximum distance between any pair of points), then the space of possibilities
can be covered by a finite number of finite size balls.  These restrictions can be motivated physically and lead to
a very general argument that there can only be a finite number of quasi-realistic string vacua;\cite{Acharya:2006zw} 
however this does not yet lead to any useful estimate of their number.

Given the topological choices of manifolds, bundles, branes and the like, one can often use algebraic geometry to
form a fairly detailed picture of a moduli space of compactifications with unbroken $N=1$ supersymmetry.
Various physical constraints, of which the simplest is the absence of long range ``fifth force'' corrections to general
relativity, imply that the scalar fields corresponding to these moduli must gain masses.  
To a large extent, this so-called ``moduli stabilization'' problem can be solved by giving the scalars
supersymmetric masses.  For example, background flux in the extra dimensions can lead to a nontrivial
superpotential depending on the moduli with many supersymmetric vacua.\cite{Giddings:2001yu}  The
many choices of flux also make the anthropic solution of the cosmological constant problem easy to 
realize.\cite{Bousso:2000xa}
Moduli stabilization also determines the distribution of vacua in the moduli space, and thus the distribution
of couplings and masses in the low energy effective theory.
One can make detailed statistical analyses of this distribution,
which incorporate and improve the traditional discussion of naturalness of couplings.\cite{Douglas:2006es}

While supersymmetric effects lift many neutral scalars, it is not at all clear that it generically lifts
all of them, satisfying bounds like \eq{moduli-bound} before taking supersymmetry breaking into account.
Explicit constructions such as that of Ref. \refcite{Kachru:2003aw} 
are usually left with one or more light scalars, and as we discussed earlier one can argue that this is 
generic.\cite{Denef:2004cf,Acharya:2010af}

Another important point which is manifest in the flux sector is what I call the ``broken symmetry paradox.''
Simply stated, it is that in a landscape, symmetry is heavily disfavored.
One can already see this in chemistry -- while the Schr\"odinger equation admits $SO(3)$ rotational symmetry,
and this is very important for the structure of atomic and molecular orbitals, once one shifts the emphasis to
studying molecules, this symmetry does not play much role.  While
a few molecules do preserve an $SO(2)$ or discrete subgroup, the resulting symmetry relations rarely have
qualitatively important consequences beyond a few level degeneracies, and it is not at all true that molecules
with symmetry are more abundant or favored in any way in chemical reactions.

It was shown in Ref. \refcite{Dine:2005gz} that discrete R symmetries are heavily disfavored in flux compactification, and
the character of the argument is fairly general.  Suppose we want vacua with a $Z_N$ symmetry; then it is
plausible that of the various parameters of some class of vacua including a symmetric point, that order $1/N$
of them will transform trivially, and order $1/N$ will each transform in one of the $N-1$ nontrivial representations.
But, since the number of vacua is exponential in the number of parameters, symmetry is extremely disfavored.
While one can imagine dynamical arguments that would favor symmetry, since these tend to operate only near
the symmetric point, it is hard to see them changing the conclusion.

One virtue of this observation is that it helps explain away the gap between the many hundreds or thousands of fields
of a typical string compactification (especially, the ones with enough vacua to solve the c.c. problem) and the
smaller number in the Standard Model, as symmetry breaking will get rid of nonabelian gauge groups and generally
lift fields.  But it is very different from the usual particle physics intuition.

\section{Eternal inflation and the master vacuum}

The wealth of disparate possibilities coming out of string compactification 
combined with the relative poverty of the data seem to force us to bring in extra structure and constraints
to help solve the vacuum selection problem and test the theory.  This will probably remain true even were we to
discover many new particles at LHC.

A good source of extra structure is cosmology, both because there is data there, and because some of the
key particle physics questions (such as low energy supersymmetry) can have cosmological consequences (such
as WIMP dark matter).  In addition to these more specific hints, as we discussed in the introduction, we have
real world examples of landscapes and we know there that the dynamics which forms metastable configurations
plays an absolutely essential role in preferring some configurations over others.  It is entirely reasonable to
expect the same here.  

A very worrying point is that the dynamics of chemistry, and even big bang nucleosynthesis, is highly nonlinear
and depends crucially on small energy differences.  The problem of deducing abundances {\it ab initio}, without
experimental data, is completely intractable.  While this might be true of the string landscape as well, in fact the
most popular scenario appears to be much simpler to analyze, as the central equations are linear.

This is the idea of eternal inflation, reviewed in Refs.~\refcite{Guth:2007ng,Linde:2007zz,Vilenkin:2011dd}
and elsewhere.  There is a good deal of current work on bringing this into string theory.  While I am not an
expert, this seems to have two main thrusts.  One is to find microscopic models of inflation or, even better,
a gauge dual to inflation analogous to AdS/CFT.  The other is to try to make the framework well enough
defined to be able to make predictions, by deriving a measure factor on the set of vacua.  We will simply cite
Ref. \refcite{Freivogel:2011eg} for a review of the status of this field and move to discussing the concrete
prescription we already quoted in the introduction,\cite{SchwartzPerlov:2006hi} 
which we call the ``master vacuum'' prescription:

\begin{quote}
The measure factor is overwhelmingly dominated by the longest lived metastable de Sitter vacuum.
For other vacua, it is given by the tunnelling rate from this ``master'' vacuum,
which to a good approximation is that of the single fastest chain of tunnelling events.
\end{quote}

Once we have convinced ourselves of this,
evidently the next order of business is to find the master vacuum.  For some measure prescriptions, this would be
an absolutely hopeless task.  For example, suppose we needed to find the metastable de Sitter vacuum with
the smallest positive cosmological constant.  By arguments from computational complexity theory,\cite{Denef:2006ad}
this problem is intractable, even for a computer the size of the universe!

The problem of finding the longest lived vacuum in this prescription
could be much easier.  A large and probably dominant
factor controlling the tunnelling rate out of a metastable vacuum is the scale of supersymmetry 
breaking.\cite{Ceresole:2006iq,Dine:2009tv}
The intuitive reason is simply that supersymmetric vacua are generally stable, by BPS arguments.
Thus, a reasonable guess is that the master vacuum is some flux sector in a vacuum with the smallest $\Lambda_{SUSY}$.
The actual positive cosmological constant is less important, both because this factor cancels out of tunnelling rates
in the analysis of the measure factor, and because there are so many choices in the flux sector available to adjust it.  
The relation to $\Lambda_{SUSY}$ also makes it very plausible that the master vacuum is {\it not} anthropically allowed.

The question of how to get small $\Lambda_{SUSY}$ deserves detailed study, but it is a very reasonable guess
that this will be achieved by taking the topology of the extra dimensions to be as complicated as possible, and
even more specifically by an extra dimensional manifold with the largest possible Euler number $\chi$.  
Of course it is intuitively reasonable that complexity allows for more possibilities and thus more extreme 
parameter values, but there is a more specific argument which we will now explain.

The first observation is that $\Lambda_{SUSY}$
is a sum of positive terms (the sum in quadrature of $D$ and $F$ breaking terms) and thus cannot receive cancellations, so
one is simply trying to make the individual $D$ and $F$ terms small.  If we imagine doing this by dynamical
supersymmetry breaking driven by an exponentially small nonperturbative effect, then the
problem is to realize a supersymmetry breaking gauge theory with the
smallest possible coupling $g^2 N$ at the fundamental scale.
This coupling is determined by moduli stabilization, and is typically related to ratios of coefficients in the
effective potential.  These coefficients can be geometric (intersection numbers, numbers of curves, etc.) or
set by quantized fluxes.  To obtain a small gauge coupling, we want these coefficients to be large.

In both cases, the typical size of the coefficients is controlled by the topology of the extra dimensions.
For example, the maximum value of a flux is determined by a tadpole or topological constraint, which for
F theory on a Calabi-Yau fourfold is
\be
\eta_{ij} N^i N^j + N_{D3} = \frac{1}{24}\chi .
\ee
Here the $N^i$ are integrally quantized values of the four-form flux, $\eta_{ij}$ is a symmetric unimodular
intersection form, and $N_{D3}$ is the number of D3-branes sitting at points in the extra dimensions.  The fluxes
$N^i$ are maximized by taking $\chi$ large and
$N_{D3}$ small, allowing large ratios of fluxes.  Although the other geometric quantities are much more
complicated to discuss, it is reasonable to expect similar relations.

Thus, we might look for the master vacuum as an F theory compactification on the fourfold with maximal $\chi$,
which (as far as I know) is the hypersurface in weighted projective space given in Ref.~\refcite{Klemm:1996ts}
with $\chi = 24\cdot 75852$.  This compactification also allows a very large enhanced gauge group with
rank 60740, including 1276 $E_8$ factors.\cite{Candelas:1997eh}
 
With this large number of cycles, the number of similar vacua obtained by varying fluxes and other choices
should be so large, $10^{10000}$ or even more, that the nearby vacua which solve the c.c. problem will be similar, 
answering the question of predictivity raised in the introduction.  But
the complexity of this compactification suggests that it might not be easy to find the precise moduli and fluxes
leading to the master vacuum.  Before doing this, we need to refine the measure factor prescription, for the
following reason.  As stated, it assumes there is a unique longest lived vacuum.  Now it is true that supersymmetry 
breaking will generate a potential on the moduli space so that de Sitter vacua will be isolated, but with this very 
small $\Lambda_{SUSY}$ these potential barriers will be incredibly small.  At the very least, one expects the tunnelling
rates to other vacua on (what was) the moduli space to be large.  It might be a better approximation
to regard the ``master vacuum'' as a distribution on this moduli space given by a simple probability measure,
perhaps uniform or perhaps a vacuum counting measure as in Ref.~\refcite{Douglas:2006es}.

The interesting tunnelling events, towards anthropically allowed vacua, would be those which increase
the scale of supersymmetry breaking.  One might imagine that supersymmetry breaking will be associated
with a single matter sector\footnote{
A recent paper on such sectors is Ref. \refcite{Simic:2010nv}.}
({\it i.e.} a minimal set $S$ of gauge groups such that no matter is charged under
both a group in $S$ and a group not in $S$) and that these tunnelling events will affect only this sector.
But since the masses of charged matter depend on moduli, in parts of the moduli space where additional matter
becomes light, one could get tunnelling events which affect other sectors as well.  We will suggest a more
intuitive picture of this dynamics in the next section.

Much is unclear about this picture.  One very basic assumption is that we can think of the cosmological dynamics
using a $4+k$-dimensional split, though of course space-time can be much more complicated.  
Better justification of this point would require a better understanding of inflation in string compactification.
If this can only be realized granting such a split (as it appears at present), this would be a justification; if not, not.
Another question is that since there are supersymmetric transitions between compactifications
with different topology, one should not even take for granted that the master vacuum is concentrated on
a single topology, though this seems plausible because such transitions change the fluxes and tadpole 
conditions.\cite{Kachru:2002gs}

\section{From hyperchemistry to phenomenology}

Granting that the dynamics of eternal inflation and the master vacuum are an important part of the vacuum selection
problem, it would be very useful to develop an intuitive picture of this dynamics.  Let us suggest such a picture based
on the assumptions stated above.  

The starting point is to think of the various structures which lead to the gauge-matter sectors relevant for
low energy physics -- groups of cycles and/or intersecting branes -- as objects which can move in the extra
dimensions.  The idea is that we are trying to describe a distribution on a pseudo-moduli space of nearly supersymmetric
vacua, and the moduli correspond to sizes of cycles, positions of branes, and the like.  Of course, the background
space in which they move will not be Euclidean or indeed any fixed geometry, and a really good picture must also
take into account this geometrical freedom.  But, with this in mind, a picture of objects moving in a fixed six dimensional
space can be our initial picture.\footnote{
Although F theory postulates a fourfold, {\it i.e.} an eight real dimensional space, two of these dimensions are
a mathematical device used to represent a varying dilaton-axion field.  The actual extra dimensions are six dimensional.
}

Next, the most important dynamics which could influence the tunnelling rates is the possibility that, as the moduli
vary, new light fields come down in mass, perhaps coupling what were previously disjoint matter sectors.  In the brane
picture, this will happen when groups of branes come close together.  Again, in the most general case, this can happen
in other ways, such as by varying Wilson lines, but let us start with the simplest case to picture.

The dynamics is thus one of structured objects (groups of cycles and intersecting branes) moving about in the
extra dimensions, and perhaps interacting when they come near each other, a sort of chemistry of the
extra dimensions.  By analogy with the familiar word `hyperspace', we might call this `hyperchemistry'.
As in chemistry, while the structures and their possible interactions are largely governed by symmetry (here the 
representation theory of supersymmetry), questions of stability and rates are more complicated to determine, though
hopefully not intractable.
 
The basic objects or molecules of hyperchemistry
are ``clusters'' of branes and cycles which intersect topologically.  These
translate into chiral gauge theories in the low energy effective theory.
Two groups of branes and cycles which do not intersect topologically
are in different clusters; these can
interact gravitationally, at long range, or by having vector-like matter become light, at short range. 

Although the nature and distribution of the clusters is not known in four dimensions, it was recently
worked out for F theory compactifications to six dimensions with eight supercharges \cite{Morrison:2012td}.
It turns out that the minimal clusters give rise to certain preferred gauge theories with matter which cannot be Higgsed,
for example $SU(2)\times SO(7)\times SU(2)$ with half-hypermultiplets in the $(2,8,1)\oplus (1,8,2)$, or
$E_8$ with no matter.  Thus a Calabi-Yau with many cycles will give rise to a low energy theory with many clusters.
A similar picture (though with different clusters) 
is expected to be true for compactifications to four dimensions as well.\footnote{D. Morrison, private communication.}

As a simple picture of the dynamics, we can imagine the clusters moving around in the extra dimensions,
occasionally undergoing transitions (tunnelling events) which change their inner structure.  Thus, we have
a fixed set of chiral gauge theories, loosely coupled to each other through bulk gravitational interactions.
Occasionally, two clusters will collide, leading to vector-like matter becoming light.  This enables further
transitions such as Higgs-Coulomb or the more complicated extremal transitions in the literature.

To some extent, the details of the extra dimensional bulk geometry would not be central to this picture; one
could get away with knowing the relative distances and orientations between each pair of clusters.  Our
previous simplifying assumption that the clusters are moving in a fixed extra dimensional geometry would
imply many constraints on these parameters, which to some extent would be relaxed by allowing the extra
dimensional geometry to vary as well.  In this way our picture could accomodate all of the relevant configurations.

Granting this picture, how might the master vacuum tunnel to an anthropically allowed vacuum?  
Now the Standard Model is a chiral gauge theory, and we know various ways to make it up out of branes and
cycles, in other words as a cluster.  It is a cluster of moderate complexity, which within F theory can be obtained 
by resolving singularities of a sort which appear naturally in fourfolds.  Thus, it is natural to imagine that
such clusters are already present in the master vacuum.  On the other hand, the master vacuum has an extremely
small supersymmetry breaking scale, probably due to dynamics in a single cluster,
with no reason to have large couplings to the Standard Model cluster.

Thus, the simplest dynamics which could create an anthropically allowed vacuum involves two steps -- 
the supersymmetry breaking cluster is modified to produce a larger scale of supersymmetry breaking,
and its interactions with the Standard Model cluster are enhanced to produce the observed supersymmetry breaking.
The first step is the one which should answer questions about the underlying scale $\Lambda_{SUSY}$
of supersymmetry breaking, while the second will determine its mediation to the observable sector.

Regarding the first, it is reasonable to expect some high scale vacua stabilized by tuned
structure in the potential as in Ref. \refcite{Denef:2004cf}, with number growing as $\Lambda_{SUSY}^{12}$ for reasons
explained there.  The number of these compared to low scale vacua with $\Lambda_{SUSY}$ exponentially small
is not yet clear.  However, granting that the master vacuum must be one with extremely small $\Lambda_{SUSY}$,
it is already a low scale vacuum, and thus the transition of the first step can easily be one which produces a
low scale vacuum, perhaps by varying a single flux and thus the gauge coupling appearing in the exponential.
Even if high scale vacua can also be produced in comparable numbers, their disadvantage in solving the hierarchy problem
will remain.  A possible loophole would be if the mediation to the Standard Model was somehow suppressed, which
seems unlikely as we argue shortly.

As for the origin of the Standard Model,
these pictures suggest that it would be realized by a single matter sector in a
localized region of the extra dimensions, in other words a local model.  This is not because it must make sense
in the decoupling limit, but rather because this is the most likely way for it to be produced by cosmological
dynamics.  Furthermore, there is no reason that the supersymmetry breaking sector must be near the Standard
Model sector or share mattter with it.  This suggests that supersymmetry breaking is generically mediated 
by supergravity interactions.

The generic estimate for scalar masses in supergravity mediation is $M_0 \sim F/M_{pl}$.   This might be smaller 
if the two sectors were ``far apart'' in the extra dimensions, but there is no known dynamics that would favor this.
As we discussed in section 2, other proposals for how this could be smaller such as sequestering are not
presently believed to be generic in string theory.  On the other hand,
it is possible for the supersymmetry breaking cluster and the Standard
Model cluster to approach very closely so that the mediation is larger.  In fact they must be closer than the
string scale and thus (from brane model intuition)
they will be coupled by vector-like matter, leading to a gauge mediation scenario.
While this is possible, since it is continuously connected to the gravitational mediation scenario, distinguished
only by varying moduli, it requires additional tuning compared to gravitational mediation.

The upshot is that gravity mediation with $M_0 \sim F/M_{pl}$ seems favored,
unless there is some reason that more of the alternative models satisfy the anthropic constraints.
This question deserves close examination by those more expert in the field than myself, but
I know of no major advantage in this regard.
Indeed, one might expect gauge mediation to lead to small $M_{3/2}$ and a cosmological moduli problem.
The picture also suggests that the $F$ terms are of the type giving rise to small gaugino
masses, since they arise in a hidden matter sector.
 
We now recall the beyond the Standard Model part of our argument.  This was to compare what seem to
be the two likeliest candidate
solutions of the hierarchy problem, namely the natural supersymmetric scenario, and the scenario
with $\Lambda_{SUSY} \sim 30-100 \TeV$ and then an additional fine tuning.  The claim is then
that the additional $10^{-5}$ or so of fine tuning gained by naturalness is more than lost by the difficulty of
solving the cosmological moduli problem, as well as meeting the other anthropic constraints, which are
much stronger in the far more complicated natural supersymmetric theories.  While this claim is hard to
argue in the absence of any knowledge about higher energy physics, if we believe we know the right
class of theories to look at on top-down grounds, we can argue it.  The class of extensions of the Standard
Model which can be realized as local models in string/M theory, interacting with a supersymmetry
breaking sector, is probably narrow enough to allow evaluating the bottom-up argument,
and making it quantitative.  As it happens, for the question
of whether we will see gauginos at LHC, it makes a great difference whether we expect 
$\Lambda_{SUSY} \sim 30 \TeV$ or $100 \TeV$ and whether F terms couple to observable gauge couplings,
and it would be great if the arguments could reach that
level of detail.

To summarize the overall picture at this point, it is that we have three sources of information about
how string/M theory could describe our universe.  Traditional particle phenomenology and astroparticle
physics are of course bottom-up and motivate model building within broad frameworks such as quantum
field theory and effective Lagrangians.  Another source is top-down, the study of compactifications and their
predictions for ``physics'' broadly construed.  The results can be summarized in effective Lagrangians, tunnelling
rates between vacua and the like, and statistical summaries of this information for large sets of vacua.
This is a ``mathematical'' definition of the landscape which could in principle
be developed {\it ab initio}, accepting only the most minimal real world input.  Finally, there is the  dynamics of 
early cosmology, by which the various vacua constructed in the top-down approach are created.  This subject is
still in its infancy -- although we have pictures such as eternal inflation which might work, the details are not yet
well understood, and there are variations and competing pictures yet to be explored.  Simplified pictures such
as hyperchemistry could help us to think physically about this dynamics.

Unless the data improves dramatically, it seems to us that all three sources must be combined to make
real predictions from string/M theory.  One must understand the set of vacua or at least those near the
master vacuum.  One must understand the dynamics of early cosmology and presumably tunnelling rates
between vacua.  In general these problems will have little or nothing to do with
either Standard Model or beyond the Standard Model physics, because the relevant dynamics is at completely
different energy and time scales.  The other part is anthropic, but given the vagueness and difficulty of working
with the anthropic principle it is probably better to simply call it ``bottom-up'' and require that we match
some or all of the data to hand.  The main difference with the existing paradigm in phenomenology is that 
we can use the top-down and early cosmology information to make a well motivated definition of naturalness,
so that if reproducing the data requires postulating an unnatural vacuum, then we have evidence against the theory.
All this is a long range project, but I think we are at the point where we can begin to work on it.

\smallskip
\noindent
{\it Acknowledgements}

I thank Bobby Acharya, Raphael Bousso, Frederik Denef,
Michael Dine, Dave Morrison,
Gordy Kane, Patrick Meade, Gary Shiu, Steve Shenker 
and Lenny Susskind for discussions and comments on the manuscript.
This research was supported in part by DOE grant DE-FG02-92ER40697.

\end{document}